\documentclass[twocolumn,showpacs,prd]{revtex4}
\usepackage{mathrsfs,bm,multirow}
\usepackage{longtable,lscape}
\usepackage{txfonts}
\usepackage{amssymb}
\usepackage{indentfirst}
\usepackage{graphicx,,booktabs}
\usepackage{color}
\usepackage{amssymb}

\usepackage[dvipdfm,  %pdftex,pdflatex
           colorlinks, %
            citecolor=blue
       ]{hyperref}
\begin{document}

\title{Towards two-body strong decay behavior of higher $\rho$ and $\rho_3$ mesons}
\author{Li-Ping He$^{1,2}$}\email{help08@lzu.edu.cn}
\author{Xiao Wang$^{1,2}$}\email{xiaowang2011@lzu.edu.cn}
\author{Xiang Liu$^{1,2}$\footnote{Corresponding author}}\email{xiangliu@lzu.edu.cn}
\affiliation{
$^1$School of Physical Science and Technology, Lanzhou University, Lanzhou 730000, China\\
$^2$Research Center for Hadron and CSR Physics, Lanzhou
University $\&$ Institute of Modern Physics of CAS, Lanzhou 730000,
China}

\begin{abstract}

In this work, we systematically study the two-body strong decay of the $\rho/\rho_3$ states, which are observed and grouped into the $\rho/\rho_3$ meson family.
By performing the phenomenological analysis, the underlying properties of these states are obtained and tested. What is more important is that abundant information of their two-body strong decays is predicted, which will be helpful to further and experimentally study these states.

\pacs{14.40.Be, 12.38.Lg, 13.25.Jx}
\end{abstract}

\maketitle

\section{introduction}\label{SS1}

There is abundant information on $\rho/\rho_3$ states collected in Particle Data Group (PDG)~\cite{Beringer:1900zz}, which provides that their
spin-parity $J^P$ could be $1^-/3^-$ and all of them are isovector. In Table~\ref{exp-rho}, we briefly review the resonance parameters of the observed $\rho/\rho_3$ states. As the total angular momentum $J$ increases, the number of these states decreases.

The experimental status of these states stimulates our interest in revealing their underlying structures, since at present the properties of $\rho/\rho_3$ are still in chaos.
First of all, we need to examine whether these $\rho/\rho_3$ can be categorized into the conventional meson family. Besides the study of mass spectrum, their Okubo-Zweig-Iizuka (OZI) allowed two-body strong decay behaviors can reflect important information on their structures. Thus, in this work we pay more efforts to systematically calculate the OZI-allowed strong decays of $\rho/\rho_3$, where the quark pair creation (QPC) model will be applied to calculation. Before carrying out calculation, we need to determine the corresponding radial, orbital and spin quantum numbers to these $\rho/\rho_3$, where we can refer to the analysis of mass spectrum, which will be summarized in the following section. By comparing our results with the experimental data, the meson assignment to these observed $\rho/\rho_3$ should be examined. Additionally, our obtained OZI-allowed two-body strong decay behaviors will provide valuable information of further experimental study on $\rho/\rho_3$.

As mentioned above, this phenomenological study on $\rho/\rho_3$ can be applied to distinguish their possible meson assignments. In addition, owing to this work, we can learn how a state has not a suitable interpretation as a conventional meson state. Thus, our study may provide important judgment whether these studies are relevant to exotic hadron configuration or new novel mechanism.

This paper is organized as follows. After introduction, we briefly review the present research status of these $\rho/\rho_3$. In Sec.~\ref{SS3}, we discuss the possible meson assignment to these states using the mass spectrum analysis and introduce the QPC model. The allowed decay channels are also selected. In Sect.~\ref{SS4}, we perform the phenomenological analysis of $\rho/\rho_3$. The last section is devoted to a short summary.

\renewcommand{\arraystretch}{1.6}
\begin{table}[htb]
\caption{The experimental information of the observed $\rho/\rho_3$ states.
Here, the masses and widths (in units of MeV) are average values taken
from PDG~\cite{Beringer:1900zz}. The states omitted from the
summary table of PDG are marked by a superscript $\flat$ while these
state as further state in PDG are distinguished by a superscript
$\natural$.   \label{exp-rho}}
\begin{center}
\begin{tabular}{lll} \toprule[1pt]
\multicolumn{3}{c}{$J^P=1^-$}\\ \midrule[1pt]
State& Mass & Width  \\\midrule[1pt]
$\rho(770)$&$775.49\pm0.34$&$146.2\pm0.7$\\
$\rho(1450)$&$1465\pm25$&$400\pm60$\\
$\rho(1570)$ $^\flat$&$1570\pm36\pm62$&$144\pm75\pm43$\\
$\rho(1700)$&$1720\pm20$&$250\pm100$\\
$\rho(1900)$ $^\flat$ \cite{Aubert:2007ym}&$1909\pm17\pm25$&$48\pm17\pm2$\\
$\rho(2150)$ $^\flat$&$2149\pm17$&$359\pm40$\\
$\rho(2000)$ $^\natural$ \cite{Anisovich:2002su,Anisovich:2000ut,Anisovich:2001jf,Anisovich:2001vt}&$2000\pm30$&$260\pm45$\\
$\rho(2270)$ $^\natural$ \cite{Anisovich:2002su,Anisovich:2000ut,Anisovich:2001jf,Anisovich:2001vt}&$2265\pm40$&$325\pm80$\\
\midrule[1pt]
\multicolumn{3}{c}{$J^P=3^-$}\\\midrule[1pt]
State& Mass & Width \\\midrule[1pt]
$\rho_3(1690)$&$1688.8\pm2.1$&$161\pm10$\\
$\rho_3(1990)$ $^\flat$ \cite{Anisovich:2002su}&$1982\pm14$&$188\pm24$\\
$\rho_3(2250)$ $^\flat$ \cite{Hasan:1994he}&$\sim2232$&$\sim220$\\
\bottomrule[1pt]
\end{tabular}
\end{center}
\end{table}

\section{Review of research status}\label{SS2}

As shown in Table~\ref{exp-rho}, there are many $\rho/\rho_3$ states observed by experiments. Among these states, $\rho(770)$ \cite{Abolins:1963zz} is established to be the ground state with $n^{2S+1}L_J=1^3S_1$ with very broad full width. Thus, we will not include $\rho(770)$ when briefly reviewing the research status of the $\rho/\rho_3$ states.
In the following, we introduce the experimental and theoretical status of $\rho/\rho_3$.

%\subsection{$\rho(1450)$ and $\rho(1700)$}

$\rho(1600)$ was omitted in the 1988 edition of PDG \cite{Yost:1988ke} and replaced by $\rho(1450)$ and $\rho(1700)$, which is due to many theoretical and experimental studies \cite{Cosme:1978qe,Bacci:1980zs,Barkov:1985ac,Dolinsky:1986kj,Erkal:1985qk,Donnachie:1986pk,Donnachie:1987as,Clegg:1988yc,Fukui:1988mp,Antonelli:1988fw,
Bisello:1988hq,Castro:1988hp,Dolinsky:1991vq,Donnachie:1994cs,Achasov:1996gw,Achasov:2000hh,Achasov:2013btb,Clegg:1993mt}.

In the past decades, many efforts have been made to explain the structure of $\rho(1450)$. However, its property is still unclear at present. Although the study of the mass spectrum supports $\rho(1450)$ as a
$2^3S_1$ state \cite{Godfrey:1985xj}, the decay behavior is hard to be understood. The calculation in Ref.~\cite{Barnes:1996ff} shows that the $\pi\pi$ and $\omega\pi$ channels are dominant in the $\rho(1450)$ decays. %, which is consistent with the experimental data \cite{Clegg:1993mt}.
Using the nonlocal Nambu-Jona-Lasinio model, the calculated partial widths of $\rho(1450)\rightarrow\pi\pi$ and $\rho(1450)\rightarrow\pi\omega$ are also comparable with the experimental values \cite{Arbuzov:2011zzf,Volkov:1997dd}.
On the other hand, the theoretical decay widths of $\rho(1450)\to a_1(1260)\pi$ and $h_1(1170)\pi$ become small \cite{Barnes:1996ff}. However, the experimental result indicates that $\rho(1450)$ mainly decays into  $4\pi$ \cite{Beringer:1900zz,Clegg:1988yc,Clegg:1993mt,Abele:2001pv}. To alleviate the discrepancy between the experimental and theoretical results of the $4\pi$ channel,
$\rho(1450)$ as a mixture of $2^3S_1$ $\rho$ state and hybrid was introduced in Ref. \cite{Barnes:1996ff} since
Close {\it et al.} indicated that the vector hybrid with mass about 1.5 GeV can strongly couple with $a_1(1260)\pi$ \cite{Close:1994hc}. Other theoretical studies \cite{Close:1994hc,Donnachie:1993hm,Close:1997dj,Donnachie:1999re} also support this mixture.

Besides these two explanations for $\rho(1450)$ as mentioned above, explanation of $\rho(1450)$ as a $1^3D_1$ state was proposed in Refs. \cite{Glozman:2010he,Glozman:2011gf} using the
chiral symmetry method. If considering the mass spectrum analysis on the $\rho$ meson family, we notice that
the mass of the $1^3D_1$ $\rho$ meson should be $1600-1700$ MeV \cite{Bugg:2012yt}. This mass discrepancy cannot be ignored when explaining $\rho(1450)$ as a $1^3D_1$ $\rho$ meson.

$\rho(1700)$ is a good candidate of the $1^3D_1$ $\rho$ meson. Both the analysis of the branching ratio of $\rho(1700)\to 2\pi,\,4\pi$ \cite{Abele:2001pv} and the study of $e^+e^-\rightarrow\omega\pi^0$ via
the nonrelativistic $^3P_0$ quark model \cite{Kittimanapun:2008wg} show that $\rho(1700)$ is a $1^3D_1$ state.

%%%%%%%%%%%%%%%%%%%%%%%%%%%%%%%%%%%%%%%%%%%%%%%%%%%%%%%%%%%%%%%%%%%%%%%%%%%%%%%%%%%%%%%%%%%%%%%%%%%%%%%%%%%%%%%%%%%%%%%%%%%%%%%
%%%%%%%%%%%%%%%%%%%%%%%%%%%%%%%%%%%%%%%  \rho(1900)      %%%%%%%%%%%%%%%%%%%%%%%%%%%%%%%5% %%%%%%%%%%%%%%%%%%%%%%%%%%%%%%%%%%%%

%\subsection{$\rho(1900)$}

There are many experiments relevant to $\rho(1900)$.
The DM2 Collaboration once reported a dip around 1.9 GeV by analyzing the $e^+e^-\rightarrow6\pi$ process \cite{Castro:1988hp}. Later, the FENICE Collaboration observed a dip around 1.9 GeV in the $R$ value measurement, which can be produced by the interference of a resonance with one of these broad vector mesons \cite{Antonelli:1996xn}. In 2001, the E687 Collaboration at Fermilab found a narrow dip structure at 1.9 GeV through the $3\pi^ +3\pi^- $ diffractive photoproduction \cite{Frabetti:2001ah} . If this dip is due to a destructive interference of a resonance with a continuum background, the resonance parameters can be extracted as $m=(1.911\pm0.004\pm 0.001)$ GeV and
$\Gamma=(29\pm 11\pm4)$ MeV.  By refitting their data, the E687 Collaboration indicated that the interference effect of a narrow resonance with known vector mesons (such as a broad $\rho(1700)$) can result in a dip \cite{Frabetti:2003pw}.  In both of $e^+ e^- \rightarrow 3\pi^+ 3\pi^- $ and $e^+ e^- \rightarrow 2\pi^+2\pi^-2\pi^0$ processes, the BaBar Collaboration announced the observation of a structure around 1.9 GeV \cite{Aubert:2006jq}, which was confirmed by BaBar in the $e^+e^-\to \phi\pi^0$ process \cite{Aubert:2007ym}.  The CMD3 Collaboration observed a peak near the $p\bar{p}$ threshold, which can be identified as  $\rho(1900)$ \cite{Solodov:2011dn}. In Ref. \cite{Bugg:2012yt}, Bugg indicated that this CMD3's observation can be explained to be a $^3S_1$ state captured by the very strong $p\overline{p}$ S-wave or to be a non-resonant cusp effect.

%%%%%%%%%%%%%%%%%%%%%%%%%%%%%%%%%%%%%%%%%%%%%%%%%%%%%%%%%%%%%%%%%5%%%%%%%%%%%%%%%%%%%%%-%%%%%%%%%%%%%%%%%%%%%%%%%%%%%%%%%%%%%%
%%%%%%%%%%%%%%%%%%%%%%%%%%%%%    \rho_(2150)      %%%%%%%%%%%%%%%%%%%%%%%%%%%%%%%5% %%%%%%%%%%%%%%%%%%%%%%%%%%%%%%%%%%%%%%%%%%

%\subsection{$\rho(2150)$}

Analyzing the data of the $6\pi$ mass spectrum from the $e^+ e^-$ annihilation \cite{Bisello:1981sh} and the diffractive photoproduction \cite{Atkinson:1985yx}, Clegg and Donnachie indicated the existence of $\rho(2150)$ \cite{Clegg:1989mp}. Later, Biagini {\it et al.} \cite{Biagini:1990ze} suggested that there exists the third radial excitation of $\rho(770)$ by phenomenologically fitting the pion form factor \cite{Dubnicka:1987tz}, and gave the corresponding resonant parameters $m\simeq2150$ MeV and $\Gamma\simeq320$ MeV, which is consistent with the result in Ref. \cite{Clegg:1989mp}. In addition, the GAMS Collaboration also confirmed the observation of $\rho(2150)$ in $\pi^-p\to \omega \pi^0 n$ \cite{Alde:1994jm,Alde:1992wv}. In Refs. \cite{Anisovich:1999xm,Anisovich:2001vt,Anisovich:2000ut,Anisovich:2002su}, the Crystal Barrel data was analyzed, where a $1^{--}$ resonance with the mass 2.15 GeV can be as the evidence of $\rho(2150)$.
In 2007, BaBar observed $\rho(2150)$ %with $m=(2.15\pm0.04\pm0.05)$ GeV and $\Gamma=(0.35\pm0.04\pm0.05)$ GeV 
in the new process $e^+e^-\to \eta^\prime(958)\pi^+\pi^-$ and $f_1(1285)\pi^+\pi^-$
\cite{Aubert:2007ef}.

Godfrey and Isgur has predicted a $2^3D_1$ state with mass 2.15 GeV \cite{Godfrey:1985xj}, which can correspond to $\rho(2150)$. However, there exists another explanation to $\rho(2150)$, i.e., Anisovich {\it et al.} suggested $\rho(2150)$ to be a $4^3S_1$ state \cite{Anisovich:2000kxa},
which was confirmed in Ref. \cite{Masjuan:2012gc,Bugg:2012yt,Masjuan:2013xta}.

%%%%%%%%%%%%%%%%%%%%%%%%%%%%%%%%%%%%%%%%%%%%%%%%%%%%%%%%%%%%%%%%%%%%%%%%%%%%%%%%%%%%%%%%%%%%%%%%%%%%%%%%%%%%%%%%%%%%%%%%%%%%
%%%%%%%%%%%%%%%%%%%%%%%%%  \rho(2000)   \rho(2270)  %%%%%%%%%%%%%%%%%%%%%%%%%%%%%%%%%%%%%%%%%%%%%%%

%\subsection{$\rho(2000)$ and $\rho(2270)$}

In Table. \ref{exp-rho}, there are two more states of $1^{--}$ listed in PDG \cite{Beringer:1900zz}, which are $\rho(2000)$ and $\rho(2270)$. In the $p\bar p \rightarrow\pi\pi$ reaction, a resonance around 1988 MeV was found \cite{Hasan:1994he}. Later, Anisovich {\it et al.} obtained a $J^{PC}=1^{--}$ state at 2000 MeV in the same reaction \cite{Anisovich:2000ut}, which also appears in the $p\bar{p}\to \omega\eta\pi^0$ and $\omega\pi$ processes \cite{Anisovich:2002su,Anisovich:2001vt,Anisovich:2001jf}. $\rho(2000)$ was suggested as the radial excitation of $\rho(1700)$ \cite{Anisovich:2002su}. In Ref. \cite{Bugg:2004xu}, Bugg concluded that $\rho(2000)$ can be a mixed state with a significant $^3D_1$ component.

In the reaction $\gamma p\rightarrow\omega\pi^+\pi^-\pi^0$, a resonance at $2280\pm50$ MeV was reported by the Omega Photon Collaboration \cite{Atkinson:1985yx}. The analysis of the Crystal Barrel data indicates that
$\rho(2270)$ is important to fit the $\omega\eta\pi$ data, and can be ignored to describe the $\omega\pi$ data \cite{Anisovich:2002su}. The Regge trajectory analysis shows that $\rho(2270)$ can be a $3^3D_1$ state, i.e., the second radial excitation of $\rho(1700)$.

%%%%%%%%%%%%%%%%%%%%%%%%%%%%%%%%%%%%%%%%%%%%%%%%%%%%%%%%%%%%%%%%%%%%%%%%%%%%%%%%%%%%%%%%%%%%%%%%%%%%%%%%%%%%%%%%%%%%%%%%%%%%
%%%%%%%%%%%%%%%%%%%%%%%%%  \rho_3(1690),~~\rho_3(1990) ~~and ~~\rho_3(2250)  %%%%%%%%%%%%%%%%%%%%%%%%%%%%%%%%%%%%%%%%%%%%%%%

In PDG \cite{Beringer:1900zz}, there are three $\rho_3$ states. $\rho_3(1690)$ was first observed in Refs. \cite{Goldberg:1965zzc,Forino:1965zz}, which was once regarded as a $\pi^+\pi^-$ resonance. At present,
$\rho_3(1690)$ is established to be a $^3D_3$ state, which can decay into $2\pi$,  $K\bar K$, $K\bar K\pi$, $4\pi$, $\omega\pi $, and $\eta\pi^+\pi^-$ as shown in PDG \cite{Beringer:1900zz}. Additionally,
two more new decay modes $a_2(1320)\pi$ and $\rho\eta$ were reported in Ref. \cite{Amelin:2000nm}. {Besides the
$^3D_3$ explanation for $\rho_3(1690)$, it could be interpreted as a three-rho meson molecular state in Ref. \cite{Roca:2010tf}.}

As a $3^{--}$ state, $\rho_3(1990)$ with $m\sim2007$ MeV and $\Gamma\sim287$ MeV was observed in the $\pi\pi$ invariant mass spectrum of $p\bar{p}\to \pi\pi$ \cite{Hasan:1994he}, which was confirmed by analyzing the Crystal Barrel data \cite{Anisovich:1999xm,Anisovich:2001jf,Anisovich:2000ut}, where a
$3^{--}$ resonance exists in the $p\bar{p}\to\pi^+\pi^-,\,\omega\pi$ processes. The $\omega\eta \pi^0$ decay of $\rho_3(1990)$ was reported in Ref. \cite{Anisovich:2001vt}. In Ref. \cite{Anisovich:2002su}, a combined fit to the $\omega\pi$, $\omega\eta\pi^0$ and $\pi^-\pi^+$ data was performed, which gives the weighed mean of mass and width of $\rho_3(1990)$ as listed in Table \ref{exp-rho}.

There are many experimental papers relevant to $\rho_3(2250)$ as shown in PDG \cite{Beringer:1900zz}.
$\rho_3(2250)$ was first observed by BNL through studying the S-channel $\bar pN$ cross section \cite{Abrams:1969jm}. Later, $\rho_3(2250)$ was also found in the reactions $p\bar p\to \bar pp$ \cite{Coupland:1977zr}, $p\bar p\to \bar N N$ \cite{Cutts:1974vi}, $p\bar p\to K^+K^-$ \cite{Carter:1978ux}, and $p\bar p\to\pi\pi$ \cite{Carter:1977jx,Martin:1980ia,Martin:1980rf,Hasan:1994he}. In 2000, the VES Collaboration reported a $3^{--}$ resonance at 2290 MeV in the reaction $\pi^-p \to\eta\pi^+\pi^-n$ \cite{Amelin:2000nm}.
The analysis of the Crystal Barrel data for the $p\bar p\to\pi^-\pi^+$ \cite{Anisovich:2000ut}, $p\bar p\to \omega\eta\pi^0$ \cite{Anisovich:2001vt} and $p\bar p\to\omega\pi$ \cite{Anisovich:2001jf} reactions also requires the existence of $\rho_3(2250)$.

A plot of the Regge trajectory for the mass spectrum of the $3^{--}$ states was presented in Refs. \cite{Anisovich:2002su,Anisovich:2001jf,Anisovich:2000ut}, where $\rho_3(1990)$ and $\rho_3(2250)$ are treated as the $2^3D_3$ and $3^3D_3$ states, respectively.

%%%%%%%%%%%%%%%%%%%%%%%%%%%%%%%%%%%%%%%%%%%%%%%%%%%%%%%%%%%%%%%%%%%%%%%%%%%%%%%%%%%%%%%%%%%%%%%%%%%%%%%%%%%%%%%%%%%%%%%%%%%%%%%%
%%%%%%%%%%%%%%%%%%%%%%%%%%%   mass spectrum  %%%%%%%%%%%%%%%%%%%%%%%%%%%%%%%%%%%%%%%%%%%%%%%%%%%%%%%%%%%%%%%%%%%%%%%%%%%%%%%%%%%

\section{Two-body strong decays}\label{SS3}

Before carrying out the study of the two-body strong decay of
these $\rho/\rho_3$ states shown in Table \ref{exp-rho}, we need
to illustrate the analysis of their Regge trajectory.

\begin{figure}[htb]
\centering
\includegraphics[width=3.5in,height=2.5in]{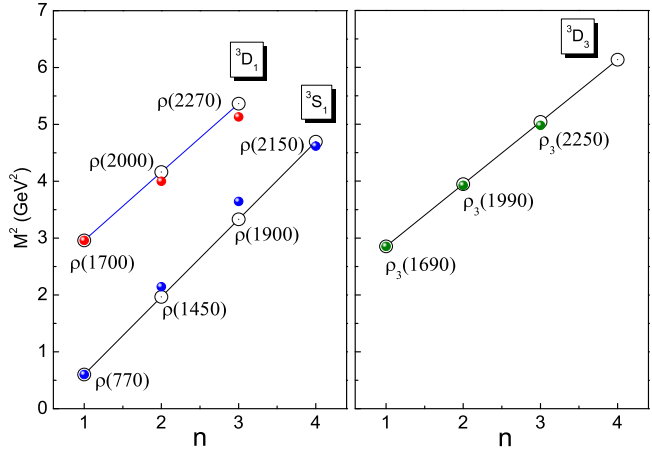}
\caption{(color online). The analysis of the Regge trajestories for the $\rho/\rho_3$
states. The trajectory slopes are 1.365 GeV$^2$, 1.203 GeV$^2$ and 1.094
GeV$^2$ for the $^3S_1$, $^3D_1$ and $^3D_3$ states, respectively. $\odot$
denotes the theoretical values, while the red, blue and green dots correspond
to the experimental data listed in Table \ref{exp-rho}.
\label{mass} }
\end{figure}

The analysis of the Regge trajectory is an effective approach
to quantitatively study meson mass spectrum. In general, there
exists an expression \cite{Chew:1962eu,Anisovich:2000kxa}
\begin{eqnarray}
M^2=M_0^2+(n-1)\mu^2,\label{rt}
\end{eqnarray}
where $M_0$ is the mass of ground state and $\mu^2$ denotes the
trajectory slope and $n$ is the radial quantum number of the
corresponding meson with mass $M$. The relation expressed by Eq.~(\ref{rt})
is roughly satisfied by $\rho/\rho_3$ states as shown
in Fig. \ref{mass}, which indicates:

\begin{enumerate}

\item{$\rho(1450)$, $\rho(1900)$ and $\rho(2150)$ are the radial excitations of $\rho(770)$. }

\item{$\rho(1700)$, $\rho(2000)$ and $\rho(2270)$ can be grouped into the $n^3D_1$ $\rho$ meson family. Among these three states, $\rho(1700)$ is the ground state while $\rho(2000)$ and $\rho(2270)$ are the first and the second radial excitations of $\rho(1700)$.}

\item{$\rho_3(1690)$, $\rho_3(1990)$ and $\rho_3(2250)$ can be as good candidates of  the $1^3D_3$, $2^3D_3$ and $3^3D_3$ states, respectively.}

\end{enumerate}

Fig. \ref{mass} only gives a rough estimate of categorizing
$\rho/\rho_3$ states into the meson
families. A further study of their two-body strong decay behaviors
can test whether the assignment shown in Fig. \ref{mass} is
reasonable. Here, the QPC model is adopted to calculate the
partial decay widths of these decays.

The QPC model was first proposed by Micu \cite{Micu:1968mk} and
further developed by the Orsay group \cite{Le Yaouanc:1972ae,Le
Yaouanc:1973xz,Le Yaouanc:1974mr,LeYaouanc:1977gm,Le Yaouanc:1977ux}.
It is has been widely adopted to study the
OZI-allowed strong decay of hadrons \cite{vanBeveren:1979bd,vanBeveren:1982qb,Bonnaz:2001aj,Roberts:1992aj,Lu:2006ry,Luo:2009wu,Blundell:1995ev,Page:1995rh,Capstick:1986bm,Capstick:1993kb,
Ackleh:1996yt,Close:2005se,Zhou:2004mw,Guo:2005cs,Zhang:2006yj,Chen:2007xf,Li:2008mza,Sun:2009tg,Liu:2009fe,
Sun:2010pg,Yu:2011ta,Wang:2012wa,Ye:2012gu}. For depicting a
quark-antiquark pair created from the vacuum, a transition
operator $\mathcal{T}$ is introduced by
\begin{eqnarray} \label{eq2}
\mathcal{T}&=&-3\gamma\sum_m \langle1m; 1-m|00\rangle \int d^3\mathbf{p}_3d^3\mathbf{p}_4\delta^3(\mathbf{p}_3+\mathbf{p}_4) \nonumber \\
           &&\times \mathcal{Y}_{1m}(\frac{\mathbf{p}_3-\mathbf{p}_4}{2})\chi^{34}_{1,-m}\phi^{34}_0\omega^{34}_0
           b^{\dagger}_{3i}(\mathbf{p}_3)d^{\dagger}_{4j}(\mathbf{p}_4).
\end{eqnarray}
Here, $\mathbf{p}_3/\mathbf{p}_4$ denotes the three momentum of quark/antiquark created from the vacuum.
Thus, the transition matrix element of the $A\to B+C$ process can
be expressed as
\begin{eqnarray}
\langle
BC|\mathcal{T}|A\rangle=\delta^3(\mathbf{P}_B+\mathbf{P}_C)\mathcal{M}^{M_{J_A}M_{J_B}M_{J_C}},\label{matrix}
\end{eqnarray}
where $\mathbf{P}_B/\mathbf{P}_C$ is the three momentum of the final state hadron
$B/C$ in the center of mass frame of the initial state $A$. In Eq.
(\ref{matrix}), $\mathcal{Y}_{\ell m}(\mathbf{p})\equiv
|\mathbf{p}|^\ell Y_{\ell m}(\theta_p,\phi_p)$ denotes the $\ell$-th
solid harmonic polynomial, $\chi^{34}_{1,-m}$ is a spin triplet
state, $i$ and $j$ are the $SU(3)$ color indices of the created
quark pairs from the vacuum. $\phi^{34}_0=(u \bar u +d \bar d +
s\bar s)/\sqrt{3}$ describes flavor singlet and
$\omega^{34}_0=\delta_{\alpha_3\alpha_4}/\sqrt{3}$
$(\alpha=1,2,3)$ corresponds to color singlet.

By the Jacob-Wick formula \cite{Jacob:1959at}, the decay amplitude
is expressed as
\begin{eqnarray}
\mathcal{M}^{JL}(A\rightarrow
BC)&=&\frac{\sqrt{2L+1}}{2J_A+1}\sum_{M_{J_B},M_{J_C}}\langle
L0;JM_{J_A}|J_A M_{J_A}\rangle\nonumber \\ \nonumber  &&\times
\langle J_B M_{J_B};J_C M_{J_C}|J M_{J_A}\rangle
\mathcal{M}^{M_{J_A}M_{J_B}M_{J_C}}.
\end{eqnarray}
Furthermore, the decay width reads as
\begin{eqnarray}
\Gamma_{A\rightarrow BC}=\pi^2\frac{|\mathbf{P}_B|}{m_A^2}\sum_{J,L}|\mathcal{M}^{JL}|^2,
\end{eqnarray}
where $m_A$ is the mass of the initial state $A$. In the concrete
calculation, the harmonic oscillator wave function
\begin{eqnarray}
\Psi_{n,\ell,m}(R,\mathbf{q})=\mathcal{R}_{n,\ell}(R,\mathbf{q})\mathcal{Y}_{\ell
m}(\mathbf{q})
\end{eqnarray}
is applied to describe the meson wave function. In the QPC model,
two parameters $R$ and $\gamma$ are introduced. Here, $R$ can be
determined by reproducing the realistic root mean square radius
which is obtained by solving the Schr\"{o}dinger equation with the
potential in Ref. \cite{Close:2005se}. {Although the $R$ values can be obtained by the above approach in principle,
these values are to be used for reference only. Thus, we illustrate the calculated partial decay widths of these $\rho$ and $\rho_3$ states in terms of parameter $R$ within a typical range of values.}
$\gamma$ is a dimensionless
constant for describing the strength of the quark pair creation.
By systematically fitting the experimental data, $\gamma=8.7$ is
obtained for $u \bar u/d \bar d$ pair creation ({see Table II in Ref. \cite{Ye:2012gu} for more details in extracting the $\gamma$ value}), while the strength
of the $s\bar s$ pair creation satisfies $\gamma=8.7/\sqrt{3}$
\cite{LeYaouanc:1977gm}.

In Table \ref{channel-rho}, the allowed two-body strong decay
channels of $\rho/\rho_3$ are listed. Using the
QPC model, we obtain the corresponding partial decay widths. In
the next section, we will compare our theoretical results with the
experimental data to perform a phenomenological analysis, which will
be helpful to further reveal the underlying properties of these
$\rho/\rho_3$ states.

\renewcommand{\arraystretch}{1.6}
\begin{table*}[htbp]
\caption{The OZI-allowed two-body decay modes of the $\rho/\rho_3$
states. Here, $\omega$, $\rho$ and $\eta^\prime$ denote
$\omega(782)$, $\rho(770)$ and $\eta'(958)$, respectively. The
allowed two-body decays are marked by $\checkmark$.
\label{channel-rho}}
\begin{center}
\begin{tabular}{c|cccccccccc} \toprule[1pt]
%\multirow{2}{c|}{$J^P=1^-$}               &\\\hline
~&$~~\pi\pi~~$    &$\pi h_1(1170)$  &$\pi\pi(1300)$   &$\pi\omega(1420)$ &$\pi\omega(1650)$   &$\rho b_1(1235)$ &$\rho f_1(1285)$  &$\omega a_1(1260)$  &$\eta \rho(1450)$  &$\rho f_1(1420)$     \\ \midrule[1pt]

$\rho(1450)$   &$\checkmark$  &$\checkmark$  &$\checkmark$  &&&&&&&\\
$\rho(1700)$   &$\checkmark$  &$\checkmark$  &$\checkmark$  &$\checkmark$ &&&&&&\\
$\rho(1900)$   &$\checkmark$  &$\checkmark$  &$\checkmark$  &$\checkmark$ &$\checkmark$ &&&&&\\
$\rho(2000)$   &$\checkmark$  &$\checkmark$  &$\checkmark$  &$\checkmark$ &$\checkmark$ &&&&&\\
$\rho(2150)$   &$\checkmark$  &$\checkmark$  &$\checkmark$  &$\checkmark$ &$\checkmark$ &$\checkmark$ &$\checkmark$ &$\checkmark$ &$\checkmark$ &\\
$\rho(2270)$   &$\checkmark$  &$\checkmark$  &$\checkmark$  &$\checkmark$ &$\checkmark$ &$\checkmark$ &$\checkmark$ &$\checkmark$ &$\checkmark$ &$\checkmark$\\
$\rho_3(1690)$ &$\checkmark$  &$\checkmark$  &$\checkmark$  &$\checkmark$ &&&&&&\\
$\rho_3(1990)$ &$\checkmark$  &$\checkmark$  &$\checkmark$  &$\checkmark$ &$\checkmark$ &&&&&\\
$\rho_3(2250)$ &$\checkmark$  &$\checkmark$  &$\checkmark$  &$\checkmark$ &$\checkmark$ &$\checkmark$&$\checkmark$&$\checkmark$ &$\checkmark$ &$\checkmark$\\
\midrule[1pt]

~&$\pi\omega$ &$\pi a_1(1260)$  &$\pi a_2(1320)$  &$\pi a_0(1450)$   &$\pi\omega_3(1670)$  &$\rho f_2(1270)$ &$\rho \eta(1295)$ &$\omega \pi(1300)$  &$\pi a_4(2040)$    &$\rho\rho(1450)$   \\\midrule[1pt]
$\rho(1450)$   &$\checkmark$  &$\checkmark$  &$\checkmark$  &&&&&&&\\
$\rho(1700)$   &$\checkmark$  &$\checkmark$  &$\checkmark$  &$\checkmark$ &&&&&&\\
$\rho(1900)$   &$\checkmark$  &$\checkmark$  &$\checkmark$  &$\checkmark$ &$\checkmark$ &&&&&\\
$\rho(2000)$   &$\checkmark$  &$\checkmark$  &$\checkmark$  &$\checkmark$ &$\checkmark$ &&&&&\\
$\rho(2150)$   &$\checkmark$  &$\checkmark$  &$\checkmark$  &$\checkmark$ &$\checkmark$ &$\checkmark$ &$\checkmark$ &$\checkmark$ &$\checkmark$ &\\
$\rho(2270)$   &$\checkmark$  &$\checkmark$  &$\checkmark$  &$\checkmark$ &$\checkmark$ &$\checkmark$ &$\checkmark$ &$\checkmark$ &$\checkmark$ &$\checkmark$\\
$\rho_3(1690)$ &$\checkmark$  &$\checkmark$  &$\checkmark$  &$\checkmark$ &&&&&&\\
$\rho_3(1990)$ &$\checkmark$  &$\checkmark$  &$\checkmark$  &$\checkmark$ &$\checkmark$ &&&&&\\
$\rho_3(2250)$ &$\checkmark$  &$\checkmark$  &$\checkmark$  &$\checkmark$ &$\checkmark$ &$\checkmark$&$\checkmark$&$\checkmark$ &$\checkmark$ &$\checkmark$\\
\midrule[1pt]

~&~~~$\eta\rho~~~$ &$\rho\rho$   &$\rho \eta'$   &$\eta b_1(1235)$  &$\pi\pi_2(1670)$  &$\pi\pi(1800)$   &$\omega a_2(1320)$       &$\eta\rho_3(1690)$  &$\rho \eta(1475)$   &$\omega a_0(1450)$   \\\midrule[1pt]
$\rho(1450)$   &$\checkmark$ &&&&&&&&&\\
$\rho(1700)$   &$\checkmark$ &$\checkmark$ &&&&&&&&\\
$\rho(1900)$   &$\checkmark$ &$\checkmark$ &$\checkmark$ &$\checkmark$ &$\checkmark$ &&&&&\\
$\rho(2000)$   &$\checkmark$ &$\checkmark$ &$\checkmark$ &$\checkmark$ &$\checkmark$ &$\checkmark$ &&&&\\
$\rho(2150)$   &$\checkmark$ &$\checkmark$ &$\checkmark$ &$\checkmark$ &$\checkmark$ &$\checkmark$ &$\checkmark$ &&& \\
$\rho(2270)$   &$\checkmark$ &$\checkmark$ &$\checkmark$ &$\checkmark$ &$\checkmark$ &$\checkmark$ &$\checkmark$ &$\checkmark$ &$\checkmark$ &$\checkmark$ \\
$\rho_3(1690)$ &$\checkmark$ &$\checkmark$ &&&&&&&&\\
$\rho_3(1990)$ &$\checkmark$ &$\checkmark$ &$\checkmark$ &$\checkmark$ &$\checkmark$ &$\checkmark$ &&&&\\
$\rho_3(2250)$ &$\checkmark$ &$\checkmark$ &$\checkmark$ &$\checkmark$ &$\checkmark$ &$\checkmark$ &$\checkmark$ &$\checkmark$ &$\checkmark$ & \\
\midrule[1pt]

~ &$K K$   &$K K^{\star}$   &$K^{\star} K^{\star}$  &$K K_1(1270)$  &$K K_1(1400)$  &$K K^{\star}(1410)$   &$K K^{\star}_2(1430)$    &$K K^{\star}(1680)$   &$K^{\star} K_1(1270)$  &$\eta' b_1(1235)$ \\\midrule[1pt]
$\rho(1450)$   &$\checkmark$ &$\checkmark$ &&&&&&&&\\
$\rho(1700)$   &$\checkmark$ &$\checkmark$ &&&&&&&&\\
$\rho(1900)$   &$\checkmark$ &$\checkmark$ &$\checkmark$ &$\checkmark$ &$\checkmark$ &&&&&\\
$\rho(2000)$   &$\checkmark$ &$\checkmark$ &$\checkmark$ &$\checkmark$ &$\checkmark$ &$\checkmark$ &$\checkmark$ &&&\\
$\rho(2150)$   &$\checkmark$ &$\checkmark$ &$\checkmark$ &$\checkmark$ &$\checkmark$ &$\checkmark$ &$\checkmark$ &&&\\
$\rho(2270)$   &$\checkmark$ &$\checkmark$ &$\checkmark$ &$\checkmark$ &$\checkmark$ &$\checkmark$ &$\checkmark$ &$\checkmark$ &$\checkmark$ &$\checkmark$\\
$\rho_3(1690)$ &$\checkmark$ &$\checkmark$ &&&&&&&&\\
$\rho_3(1990)$ &$\checkmark$ &$\checkmark$ &$\checkmark$ &$\checkmark$ &$\checkmark$ &$\checkmark$ &$\checkmark$ &&&\\
$\rho_3(2250)$ &$\checkmark$ &$\checkmark$ &$\checkmark$ &$\checkmark$ &$\checkmark$ &$\checkmark$ &$\checkmark$ &$\checkmark$ &$\checkmark$ &$\checkmark$\\  %\bottomrule[1pt]
\bottomrule[1pt]
\end{tabular}
\end{center}
\end{table*}

\section{Phenomenological analysis}\label{SS4}

\subsection{$n~^3S_1$ states}

\begin{figure*}[htbp]
\includegraphics[scale=0.5]{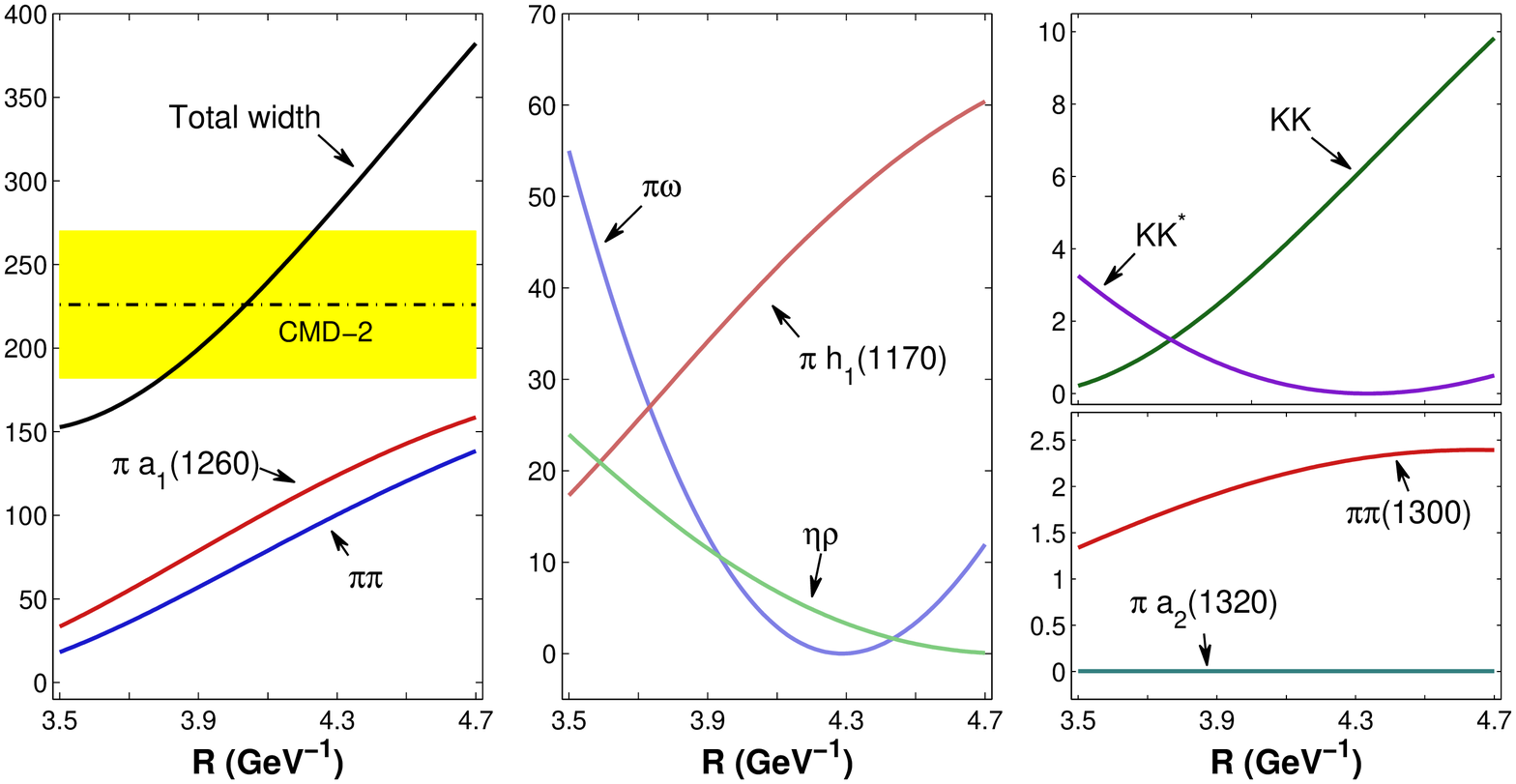}
\caption{(color online). The calculated partial and total decay widths of $\rho(1450)$ dependent
on the $R$ value. Here, the
dashed line with band is the experimental total width from Ref.
\cite{Akhmetshin:2001hm}. \label{2S}}
\end{figure*}

Assuming $\rho(1450)$ as a $2^3S_1$ isovector meson, its two-body
strong decay behavior is calculated and shown in Fig.~\ref{2S}.
Our calculation shows that $\pi\pi$, $\pi a_1(1260)$,
$\pi\omega$ and $\pi h_1(1170)$ are its dominant decay modes,
where $\pi a_1(1260)$,
$\pi\omega$ and $\pi h_1(1170)$  can contribute to the $4\pi$
final state.
%\cite{Akhmetshin:1998df,Edwards:1999fj,Bisello:1990du}.
In addition, the obtained width of $\rho(1450)\to \eta\rho$ is also
in good agreement with the data in Refs.
\cite{Beringer:1900zz,Clegg:1993mt}. The partial decay widths of
$\rho(1450)$ into $K\bar K$, $K\bar K^* +H.c.$ and
$\pi\pi(1300)$ are small in our calculation. As for
$\rho(1450)\rightarrow\pi a_2(1320)$, the decay width is tiny.
Thus, the experimental data listed in PDG \cite{Beringer:1900zz}
can be quantitatively compared with our results.
Given the information of partial decay widths, we obtain the
total width of $\rho(1450)$ by summing over all partial decay
widths. In Fig. \ref{2S}, we show the comparison of our results
with the CMD-2 data \cite{Akhmetshin:2001hm}, which indicates that
there exists a common range between our theoretical total width and
the experimental data. Additionally, the obtained total width is
also consistent with the experimental width given in Ref.
\cite{Donnachie:1986pk}, and overlaps with the measured full width
listed in Refs. \cite{Clegg:1993mt,Caso:1998tx}, which is about
310 MeV.

%The $R$ value for $\rho(1450)$ is about 4.17 GeV$^{-1}$ by the potential model \cite{Close:2005se}.
%The
%In this range,
%it couples mainly to $\pi a_1(1260)$ and $\pi \pi$, but the $\omega\pi$ is suppressed, as seen in Fig. \ref{2S}.
%When R is near 3.5 GeV$^{-1}$, $\Gamma_{\pi\omega}\thicksim55$ MeV, so the widths of $\omega\pi$ have overlap with \cite{Clegg:1993mt}.

Besides providing the information of the partial decay widths of $\rho(1450)$, in Table \ref{ratio1} several ratios $\Gamma_{\pi \pi}/\Gamma_{\pi a_1(1260)}$, $\Gamma_{\pi h_1(1170)}/\Gamma_{\pi a_1(1260)}$ and $\Gamma_{\pi a_1(1260)}/\Gamma_{Total}$ are also given, which are weakly dependent on the parameter $R$. Experimental measurement of these ratios will be a good test of the $2^3S_1$ assignment to $\rho(1450)$.

Because of the above analysis, we conclude that it is easy to explain $\rho(1450)$ as a $2^3S_1$ state, which is also supported by a recent work in Ref. \cite{Ketzer:2012vn} that claims there is no clear evidence for a hybrid state with $J^{PC}=1^{--}$.

%The decay information about $\rho(1450)$ in experiment is still scarce. In a recent paper \cite{Achasov:2012zza}, the cross section of $e^+e^-\rightarrow\omega\pi^0$ is studied, but they don't give a  specific value of the masses and widths of both $\rho(1450)$ and $\rho(1700)$. And they conclude that accurate  determination  of  the parameters of $\rho(1450)$ and $\rho(1700)$ requires higher statistics for energies above 1.4 GeV.
%The data of $\Gamma_{\pi\pi}/\Gamma_{4\pi}$ , $\Gamma_{\pi\omega}/\Gamma_{4\pi}$ and $\Gamma_{\pi\pi}/\Gamma_{\omega\pi}$ in PDG \cite{Beringer:1900zz} are contradictory
%with each other. To understand the real structure of this state, we need further decay information of $\rho(1450)$.

%\medskip$\rho(1900)$\medskip
\begin{figure*}[htbp]
\includegraphics[scale=0.55]{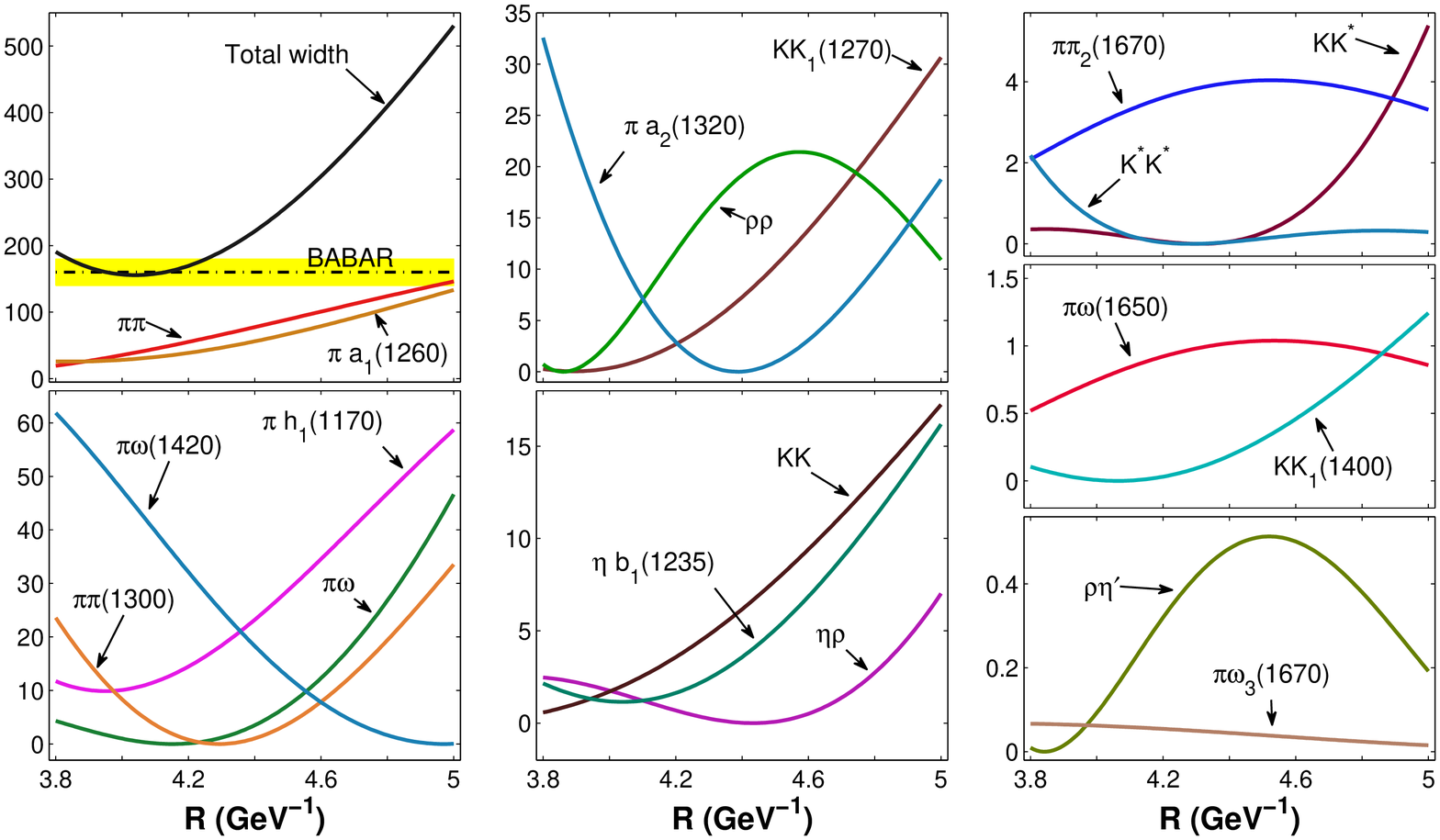}
\caption{(color online). The calculated partial and total decay widths of $\rho(1900)$ dependent
on the $R$ value. Here, we do not list the decay width of $\rho(1900)\to \pi a_0(1450)$ since this channel is tiny. The dashed line with band is the experimental total width from Babar \cite{Barnes:1996ff}. \label{3S}}
\end{figure*}

According to the Regge trajectory analysis, $\rho(1900)$ is a good candidate for a $3^3S_1$ state. At present, its resonance parameters are not yet determined experimentally, i.e., different experiments give different results as listed in PDG \cite{Beringer:1900zz}. The calculated two-body strong decays of $\rho(1900)$ are presented in Fig. \ref{3S}, where the theoretical total width overlaps with the BaBar's data \cite{Barnes:1996ff}. In addition, the main decay modes of $\rho(1900)$ are $\pi\pi$, $\pi a_1(1260)$, $\pi h_1(1170)$, $\pi\pi(1300)$, and $\pi \omega(1420)$. Thus, $\rho(1900)$ has a large $4\pi$ branching ratio and the decays into $\rho\rho$, $K\bar K$, and $\eta b_1(1235)$ are sizeable.
In Table. \ref{ratio1}, we also show several ratios of its partial decay widths. These predicted decay behaviors will be helpful to experimentally study $\rho(1900)$ in future.

%$\rho(2150)$

\begin{figure*}[htbp]
\includegraphics[scale=0.5]{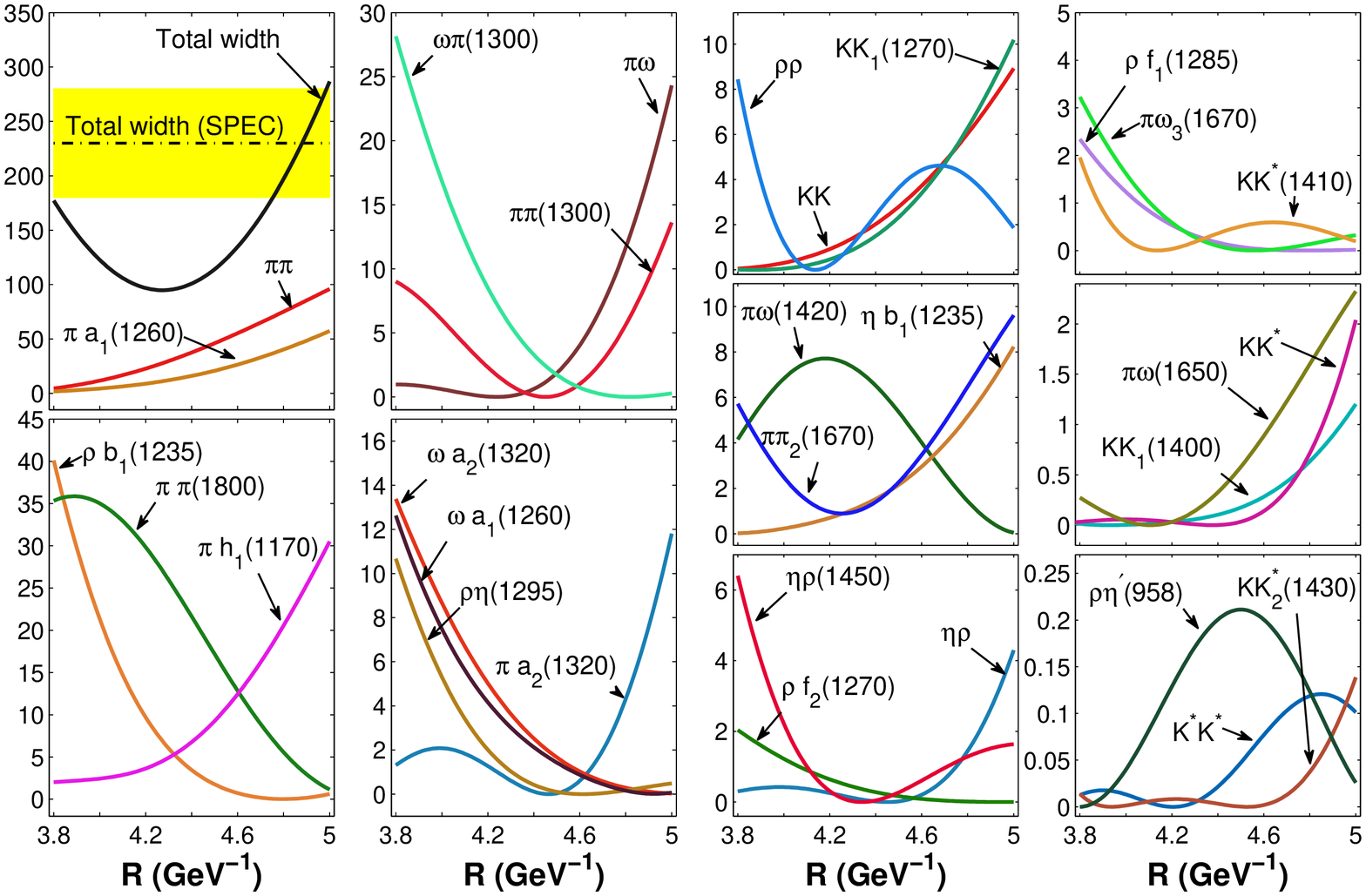}
\caption{(color online). The calculated partial and total decay widths of $\rho(2150)$ dependent
on the $R$ value. The dashed line with band is the experimental total width taken from Ref. \cite{Anisovich:2002su}.  \label{4S}}
\end{figure*}

As shown in Fig. \ref{mass}, $\rho(2150)$ can be a $4^3S_1$ state. The OZI-allowed two-body strong decay widths are listed in Fig.~\ref{4S}. The obtained total width is dependent on the $R$ value due to the node effect, where the total width is $(108-287)$ MeV corresponding to $R=(4.3-5.0)$ GeV$^{-1}$. {From PDG \cite{Beringer:1900zz}, we notice that the measured total width of $\rho(2150)$ from the $e^+e^-$ interaction is larger than that from the $p\bar{p}\to \pi\pi$ process and $S$-channel $N\bar{N}$ interaction. Here, the experimental total widths of $\rho(2150)$ are (350 \cite{Aubert:2007ef}, 389 \cite{Biagini:1990ze}, 410 \cite{Clegg:1989mp}, 310 \cite{Aubert:2007ef}) MeV, (296 \cite{Hasan:1994he}, 40 \cite{Oakden:1993he}, 250 \cite{Martin:1980ia}, 200 \cite{Martin:1980rf}) MeV, and (230 \cite{Anisovich:2002su}, 135 \cite{Coupland:1977zr}, 98 \cite{Alspector:1973er}, 85 \cite{Abrams:1969jm}) MeV corresponding to the $e^+e^-$ interaction, $p\bar{p}\to \pi\pi$ channel and $S$-channel $N\bar{N}$ process, respectively. Our calculation favors the data measured at the $p\bar{p}\to \pi\pi$ process and $S$-channel $N\bar{N}$ interaction.} For example, in Fig. \ref{4S} we compare our result of the total width with that in Ref. \cite{Anisovich:2002su} obtained by analyzing the SPEC's data, where the theoretical and experimental results overlap with each other when $R=(4.74-4.98)$ GeV$^{-1}$. The calculation of the partial decay widths shows that $\rho(2150)$ decays dominantly into $\pi\pi$, $\pi a_1(1260)$, $\pi\omega$ and $\pi h_1(1170)$. More information of other partial decay widths can be found in Fig. \ref{4S}.
We notice that $\rho(2150)$ was observed in the decay channels $\pi^+\pi^-$, $\omega\pi^0$, $\eta'\pi\pi$, $f_1(1285)\pi\pi$, $\omega\pi\eta$, $K^+K^-$ and $6\pi$ \cite{Beringer:1900zz}, which can be reasonably explained by our study. Furthermore, based on the obtained partial decay widths, we also give several ratios of some partial decay widths in Table \ref{ratio1}, which are also important to test whether $\rho(2150)$ is a $4^3S_1$ state.

\begin{table}[htbp]
\caption{The obtained ratios of the partial decay widths of the $\rho$ states discussed in Figs. \ref{2S}-\ref{1D}. {Here, we have only listed the ratios weakly dependent on $R$, which is the reason why we have not listed the ratios of $\Gamma_{\eta\rho}/\Gamma_{\omega\pi}$ and $\Gamma_{KK}/\Gamma_{\omega\pi}$ for $\rho(1450)$ that are strongly dependent on $R$.} \label{ratio1}}
%\begin{center}
\begin{tabular}{l|ccc}
%\toprule[1pt]
\toprule[1pt]
~             &$\Gamma_{\pi\pi}/\Gamma_{\pi a_1(1260)}$    &$\Gamma_{\pi h_1(1170)}/\Gamma_{\pi a_1(1260)}$          &$\Gamma_{\pi a_1(1260)}/\Gamma_{Total}$ \\\midrule[1pt]
$\rho(1450)$  &$0.545-0.873$     &$0.517-0.381$  &$0.219-0.415$\\
$\rho(1700)$  &$0.400-0.178$     &$0.728-0.696$  &$0.327-0.337$\\
$\rho(1900)$  &$0.738-1.432$     &$0.470-0.360$  &$0.129-0.262$\\
$\rho(2150)$  &$2.626-1.674$     &$1.121-0.404$  &$0.009-0.199$\\
\bottomrule[1pt]
\end{tabular}
%\end{center}
\end{table}

{The ranges of $R$ in Figs. \ref{2S}-\ref{4S} needed to reproduce the experimental total widths are $(3.79-4.23)$ GeV$^{-1}$, $(3.85-4.28)$ GeV$^{-1}$ and $(4.74-4.98)$ GeV$^{-1}$, respectively, where the experimental error is considered. These obtained ranges of $R$ also roughly reflect a regularity, i.e., the corresponding $R$ value becomes larger when the radial quantum number increases, which is consistent with the estimate of the potential model \cite{Close:2005se}.
}

\subsection{$n~ ^3D_1$ states}

The Regge trajectory analysis shows that $\rho(1700)$, $\rho(2000)$ and $\rho(2270)$
can be categorized into the $n^3D_1$ $\rho$ meson family (see Fig. \ref{mass}). In this subsection, we discuss their two-body decay behaviors.

\begin{figure*}
\includegraphics[scale=0.5]{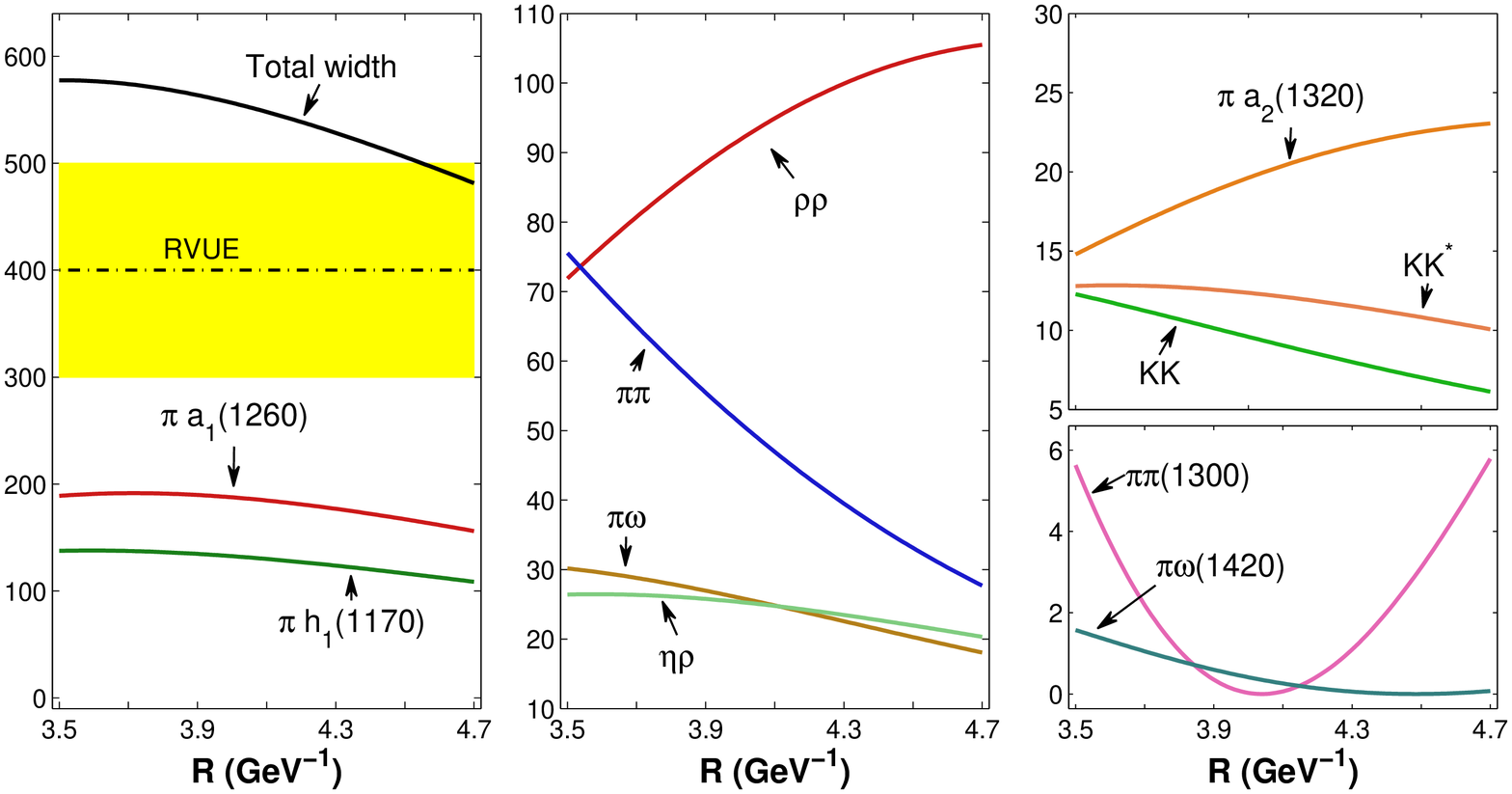}
\caption{(color online). The partial and total decay widths of $\rho(1700)$ dependent on the $R$ value. Here, we do not list $\rho(1700)\to \pi a_0(1450)$ due to its tiny decay width. The dashed line with band is the experimental total width from Ref. \cite{Clegg:1993mt}.  \label{1D}}
\end{figure*}

As the ground state of the $^3D_1$ $\rho$ meson family, $\rho(1700)$ mainly decays into $\pi a_1(1260)$ and $\pi h_1(1170)$. Of course, $\pi\pi$ and $\rho\rho$ are the important decay channels. These results are consistent with the experimental data \cite{Beringer:1900zz,Clegg:1993mt}, which naturally explains why $\rho(1700)$ can be found in its $4\pi$ and $\rho\pi\pi$ channels. However, the obtained total decay width is larger than most of experimental data listed in PDG. In Fig. \ref{1D}, we give the comparison between our result and the experimental total width from Ref. \cite{Clegg:1993mt}, where the theoretical total width can overlap the experimental result with error when $R>4.55$ GeV$^{-1}$.

In addition, we find that the decay width of $\rho(1700)\to \omega\pi$ is always smaller than that of $\rho(1700)\to \pi\pi$, which does not depend on the $R$ value. This conclusion is consistent with the results in Refs. \cite{Godfrey:1985xj,Clegg:1993mt,Kokoski:1985is}.
%However, in \cite{Clegg:1993mt}, they give a very small $\omega\pi$ modes, different from others.
The obtained decay width of
$\rho(1700)\to \pi\pi$ is comparable with the value ($39\pm4)$ MeV given in Ref. \cite{Bugg:1996ki}.
In the Godfrey-Isgur potential model \cite{Godfrey:1985xj}, the estimated decay width for $\rho(1700)\rightarrow\omega\pi$ is about 25 MeV, which well agrees with our calculation of $\rho(1700)\rightarrow\omega\pi$. For the $\rho(1700)\to \eta\rho$ decay, the calculated result is comparable with that listed in Ref. \cite{Clegg:1993mt}. In Table \ref{ratio1}, some ratios of the partial decay widths of $\rho(1700)$ are presented.

{According to PDG, as for $\rho(1700)$ the ratio $\Gamma_{\pi\pi(1300)}/\Gamma_{4\pi}$ is
0.3 while the ratio $\Gamma_{\pi a_1(1260)}/\Gamma_{4\pi}$ is 0.16 (with a large uncertainty)
 \cite{Abele:2001pv}. We need to emphasize that these ratios $\Gamma_{\pi\pi(1300)}/\Gamma_{4\pi}$ and $\Gamma_{\pi a_1(1260)}/\Gamma_{4\pi}$ listed in PDG can be changed with different considerations of fitting the experimental data (see Sec. 4.3 in Ref. \cite{Abele:2001pv} for more details). If adopting these two experimental ratios, that
would mean that the decay into $\pi\pi(1300)$ should be more likely than
into $\pi a_1(1260)$. However, we get an order of magnitude larger
decay rate into $\pi a_1(1260)$. This discrepancy should be explained when assigning $\rho(1700)$ as a $1^3D_1$ state. Introducing the exotic state explanation to $\rho(1700)$ and studying the corresponding decay behavior are an interesting topic. }

\begin{figure*}[htbp]
\includegraphics[scale=0.6]{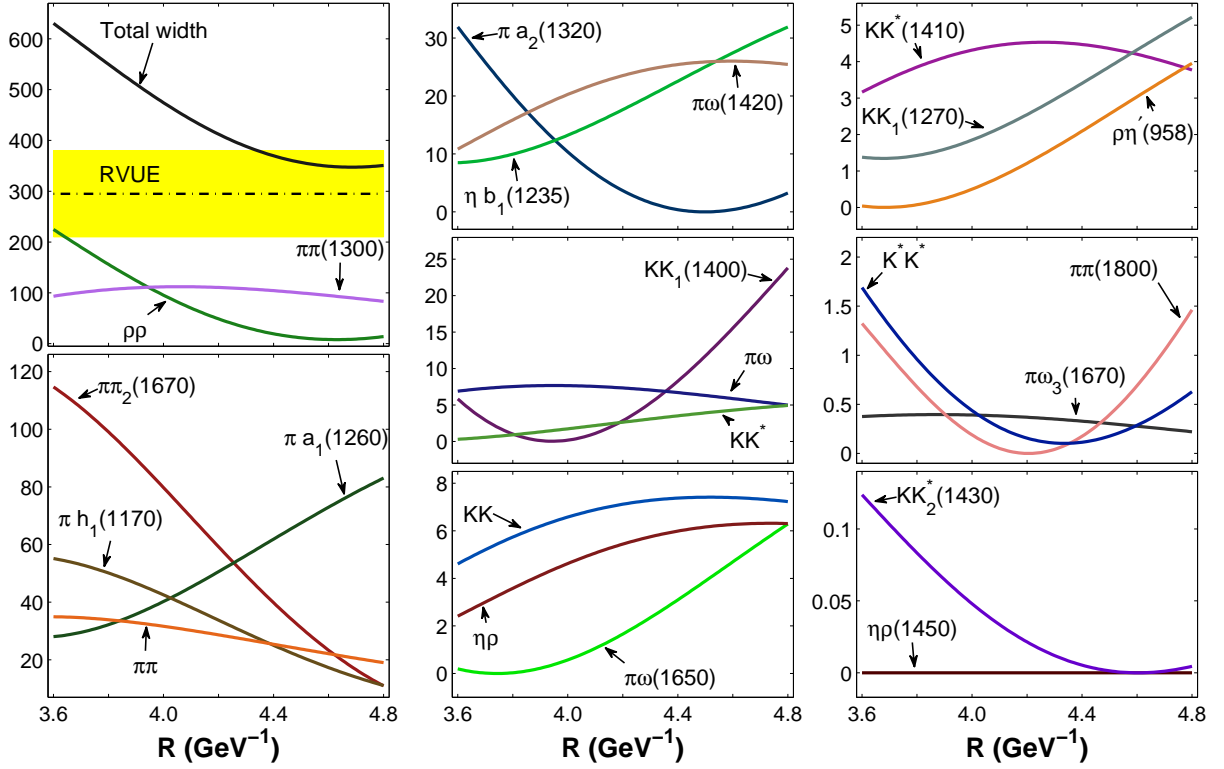}
\caption{(color online). The partial and total decay widths of $\rho(2000)$ as the $2^3D_1$ state dependent on the $R$ value. The dashed line with band is the experimental total width \cite{Anisovich:2000ut}.\label{2D}}
\end{figure*}

As the candidate of a $2^3D_1$ state, the two-body decay and total decay widths of $\rho(2000)$ are obtained in Fig. \ref{2D}. The total width can overlap with the Crystal Barrel result in Ref. \cite{Anisovich:2000ut} when $R=(4.34-4.80)$ GeV$^{-1}$. $\rho(2000)$ dominantly decays into $\pi\pi(1300)$, $\rho\rho$, $\pi\pi_2(1670)$ and $\pi a_1(1260)$. The decay channels of the $\rho(2000)$ into $\pi\pi$, $\pi h_1(1170)$, $\pi a_2(1320)$, $\pi\omega(1420)$ and $\eta b_1(1235)$ are also important.

\begin{figure*}[htbp]
\includegraphics[scale=0.52]{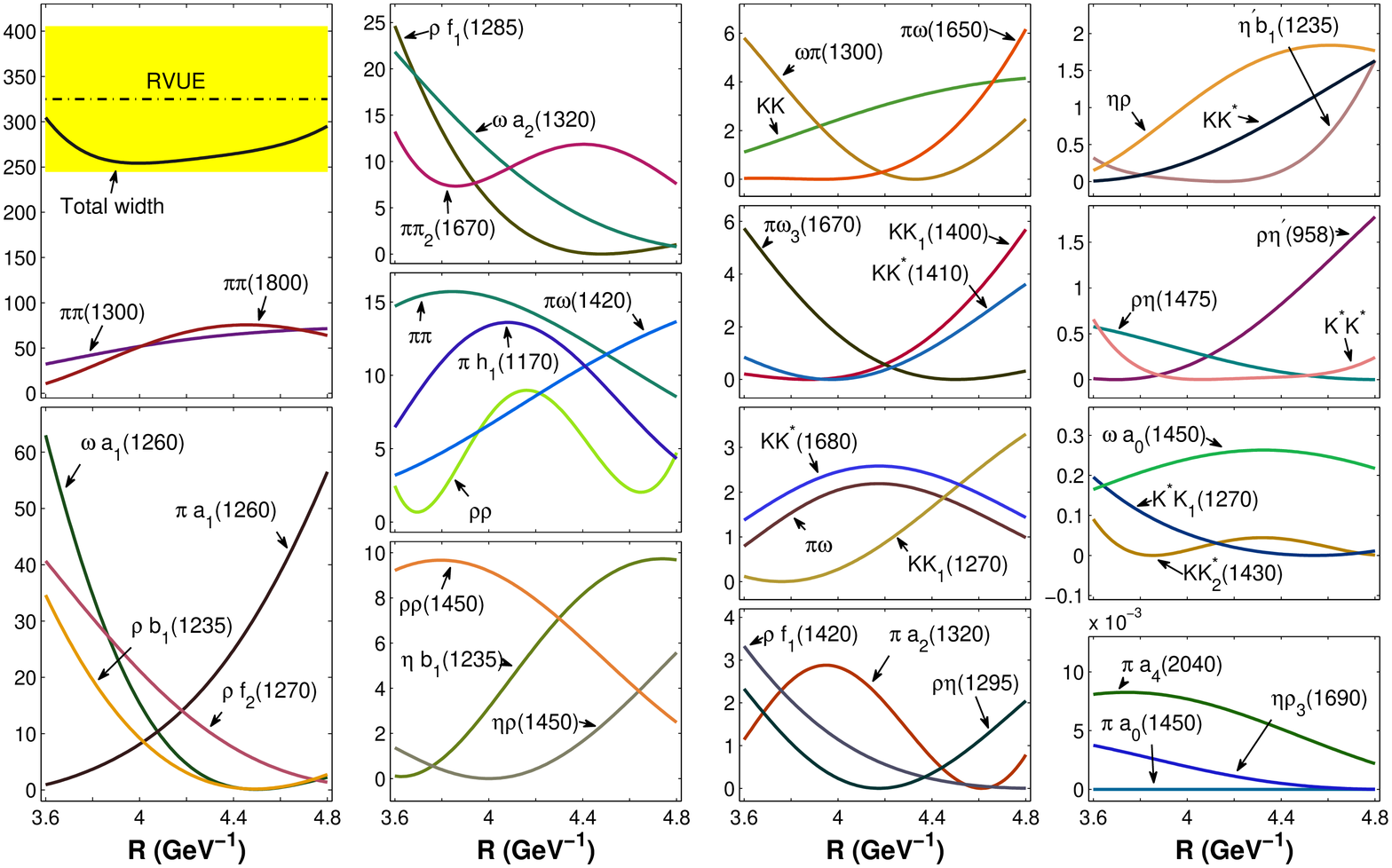}
\caption{(color online). The partial and total decay widths of $\rho(2270)$ dependent on the $R$ value. The dashed line with band is the experimental total width \cite{Anisovich:2002su}.\label{3D}}
\end{figure*}

Fig. \ref{3D} shows the decay information of $\rho(2270)$ from the calculation of the QPC model. Although more decay channels are open, $\rho(2270)$ has a smaller total decay width compared with the former two $^3D_1$ states. The obtained total decay width can overlap with the Crystal Barrel data \cite{Anisovich:2002su} as shown in Fig. \ref{3D}. The main decay modes are $\pi\pi(1300)$ and $\pi\pi(1800)$. Other important decay channels include $\pi a_1(1260)$, $\omega a_1(1260)$, $\rho f_2(1270)$, and $\rho b_1(1235)$.

%The $\pi\pi$ and $\pi h_1(1170)$ are smaller and $\pi\omega$ is negligible.

At present, experiments scarcely provide information on $\rho(2270)$. Thus, the theoretical predictions of the two-body strong decays of $\rho(2270)$ shown in Fig. \ref{3D} and Table \ref{ratio4} can provide valuable guidance to future experimental study on $\rho(2270)$.

\begin{table}[htbp]
\caption{Several calculated branching ratios of the partial decay widths of $\rho(2000)$ and $\rho(2270)$.   \label{ratio4}}
%\begin{center}
\begin{tabular}{l|ccc}
%\toprule[1pt]
\toprule[1pt]
~             &$\Gamma_{\pi a_1(1260)}/\Gamma_{\pi\pi(1300)}$    &$\Gamma_{\pi h_1(1170)}/\Gamma_{\pi\pi}$          &$\Gamma_{\pi\pi(1300)}/\Gamma_{Total}$ \\\midrule[1pt]
$\rho(2000)$  &$0.300-0.997$     &$0.634-1.714$  &$0.148-0.238$\\
$\rho(2270)$  &$0.028-0.790$     &$0.439-0.507$  &$0.108-0.253$\\
\bottomrule[1pt]
\end{tabular}
%\end{center}
\end{table}

{In Figs. \ref{1D}-\ref{3D}, we also notice that the corresponding $R$ values for reproducing the experimental data are within the allowed range.}

\subsection{$n ~^3D_3$ states}

%\medskip$\rho_3(1690)$\medskip
\begin{figure*}[htb]
\includegraphics[scale=0.6]{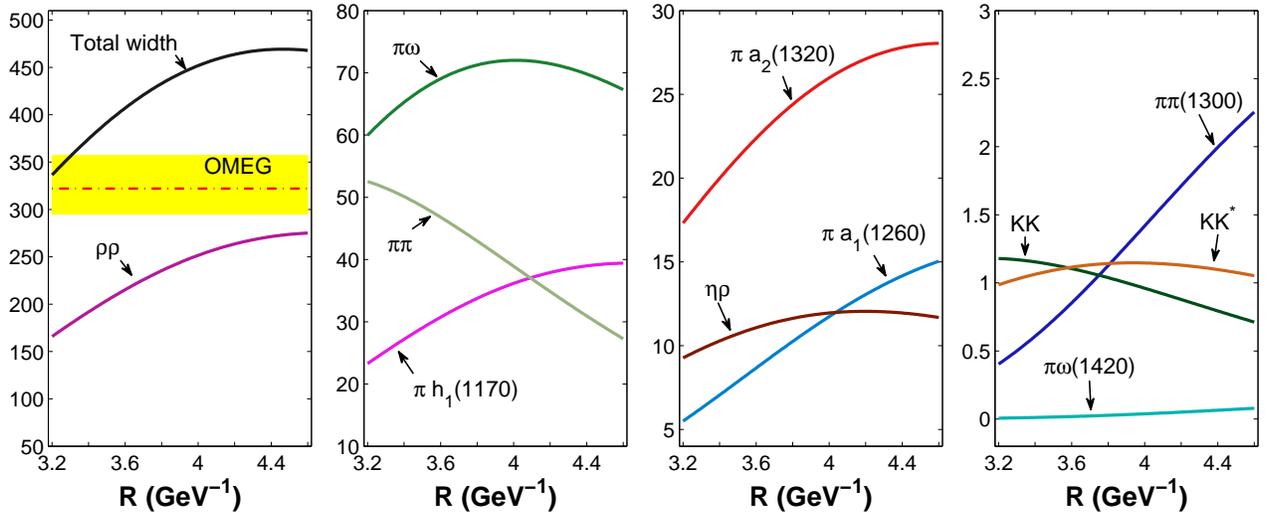}
\caption{(color online). The partial and total decay widths of $\rho_3(1690)$ dependent of the $R$ value. The dashed line with band is the experimental total width \cite{Corden:1978da}.\label{1D3}}
\end{figure*}

If $\rho_3(1690)$ is a $1^3D_3$ state,
the partial decay widths are shown in Fig. \ref{1D3}, where the decay of $\rho_3(1690)$ is dominated by the $\rho\rho$ channel. The other large decay modes include $\pi\omega$, $\pi\pi$ and $\pi h_1(1170)$. The decay modes of $\pi a_2(1320)$, $\pi a_1(1260)$ and $\eta\rho$ are also sizeable.

\begin{table}[htb]
\caption{Several calculated branching ratios of $\rho_3(1690)$ and the ratio $\Gamma_{\pi a_2(1320)}/\Gamma_{\eta\rho}$. Here, we also list the corresponding experimental data in the third column. \label{ratio2}}
%\begin{center}
\begin{tabular}{c|cc}
\toprule[1pt]
Ratios                                  & This work                         & Experimental data\\\midrule[1pt]
$\Gamma_{\rho\rho}/\Gamma_{Total}$      & $(49.33-58.79)\%$ ~~~~    &  -\\
$\Gamma_{\pi\pi} /\Gamma_{Total}$       & $(5.83-15.62)\%$ ~~~~    & $(23.6\pm1.3)\%$ \cite{Beringer:1900zz}  \\
$\Gamma_{\pi\omega}/\Gamma_{Total}$     & $(14.38-17.83)\%$ ~~~~    & $(16\pm6)\%$ \cite{Beringer:1900zz}  \\
%$\Gamma_{4\pi}/\Gamma_{Total}$  &         -               & $(71.1\pm1.9)\%$ \cite{Beringer:1900zz}  \\
$\Gamma_{\pi h_1(1170)}/\Gamma_{Total}$ & $(6.92- 8.42)\%$ ~~~~     &  - \\
%$\Gamma_{\pi a_2(1320)}/\Gamma_{Total}$ & $5.15\% - 5.99\%$ ~~~~     &   \\
%$\Gamma_{\rho\eta}/\Gamma_{Total}$      & $2.76\% - 2.50\%$ ~~~~     &   \\
$\Gamma_{\pi a_2(1320)}/\Gamma_{\eta\rho}$   & $1.87-2.40$ ~~~~     & $5.5\pm2.0$ \cite{Amelin:2000nm}\\
\bottomrule[1pt]
\end{tabular}
%\end{center}
\end{table}

\begin{table}[htb]
\caption{Some typical branching ratios of $\rho_3(1690)$, $\rho_3(1990)$ and $\rho_3(2250)$.    \label{ratio3}}
%\begin{center}
\begin{tabular}{c|ccc}
\toprule[1pt]
~      &$\Gamma_{\rho\rho}/\Gamma_{Total}$  &$\Gamma_{\pi\pi}/\Gamma_{Total}$   &$\Gamma_{\pi\pi(1300)}/\Gamma_{Total}$\\ \midrule[1pt]
$\rho_3(1690)$    &~~$0.493 - 0.588~~$  &~~$0.156- 0.058~~$ &~~$0.001 - 0.005~~$  \\
$\rho_3(1990)$    &$0.250 - 0.182$  &$0.061- 0.240$  &$0.217 - 0.068$  \\
$\rho_3(2250)$    &$0.010 - 0.068$  &$0.020- 0.358$  &$0.000001 - 0.168$  \\
\bottomrule[1pt]
\end{tabular}
%\end{center}
\end{table}

In Table \ref{ratio2}, several branching ratios of $\rho_3(1690)$ and the ratio $\Gamma_{\pi a_2(1320)}/\Gamma_{\eta\rho}$ are calculated in comparison with the corresponding experimental values.
Our branching ratios of $\rho_3(1690)\to \pi\pi,\,\pi\omega$ and the ratio $\Gamma_{\pi a_2(1320)}/\Gamma_{\eta\rho}$ are comparable with the experimental results. At present, experiments reveal that $\rho_3(1690)$ dominantly decays into $4\pi$ with the branching ratio $\sim 71.1\%$ \cite{Beringer:1900zz}, which is supported by our calculation, where the final states $\pi\omega$ and $\rho\rho$ can mainly contribute to the $4\pi$ final state.

{In Fig. \ref{1D3}, we give comparison of our results with
the experimental data \cite{Corden:1978da}. If reproducing the experimental total width, the adopted $R$ value is about 3 GeV$^{-1}$, which is unreasonable. In addition, the obtained total decay width of $\rho_3(1690)$ is larger than the data in PDG \cite{Beringer:1900zz} when taking $R$ around $4$ GeV$^{-1}$ \cite{Close:2005se}. This situation shows that $\rho_3(1690)$ as a $1^3D_3$ state
seems questionable. For clarifying this point, we suggest the precise measurement of its resonance parameters in future experiments. Of course, this discrepancy mentioned above also provides a possibility of introducing the exotic state explanation to $\rho_3(1690)$. We notice that a three-$\rho$ meson molecular state was proposed in Ref. \cite{Roca:2010tf}. }

%$\rho_3(1990)$

\begin{figure*}[htbp]
\includegraphics[scale=0.6]{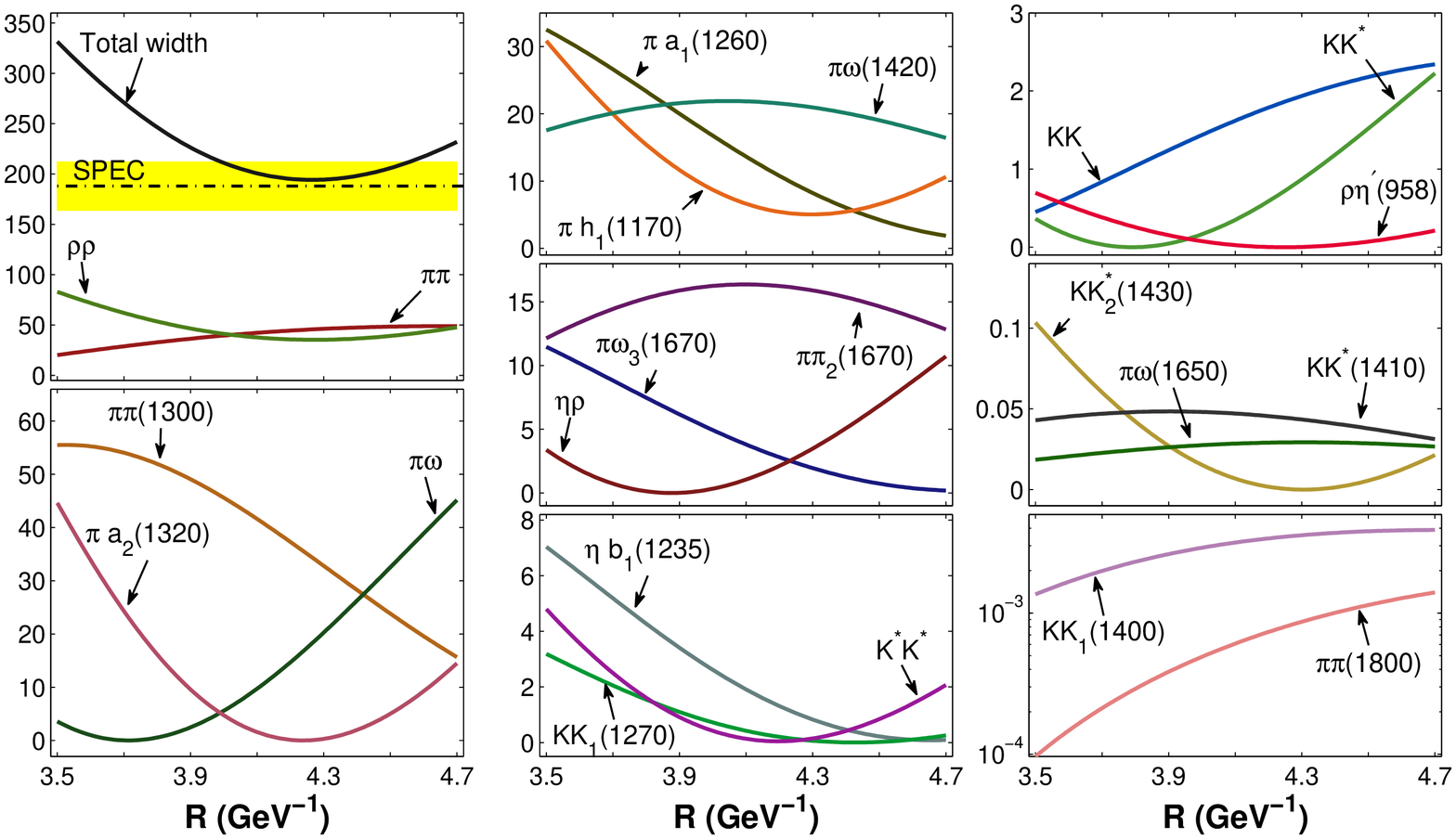}
\caption{(color online). The partial and total decay widths of $\rho_3(1990)$ dependent on the the $R$ value. Here, the $\pi a_0(1450)$ channel is not listed. The dashed line with band is the experimental total width \cite{Anisovich:2002su}. \label{2D3}}
\end{figure*}

The partial decay widths of $\rho_3(1990)$ are predicted in Fig. \ref{2D3}, where the mass of $\rho_3(1990)$ in Table \ref{exp-rho} is adopted in our calculation. $\rho_3(1990)$ mainly decays into
$\rho\rho$, $\pi\pi$, $\pi\omega$, $\pi\pi(1300)$ and $\pi\omega$. Several typical decay branching ratios of $\rho_3(1990)$ are presented in Table \ref{ratio3}. %$\pi h_1(1170)$, $\pi a_1(1260)$, $\pi a_2(1320)$, and  $\pi\omega(1420)$.
The calculated total decay width of $\rho_3(1990)$ is compatible with the experimental data \cite{Anisovich:2002su} as shown in Fig. \ref{2D3}. In addition, $\rho_3(1990)\to \pi\pi,\,\pi\omega$ were observed in the experiment \cite{Beringer:1900zz}. Our calculation shows that the decay widths of $\rho_3(1990)\to \pi\pi,\,\pi\omega$  are sizeable.

%$\rho_3(2250)$

\begin{figure*}
\includegraphics[scale=0.5]{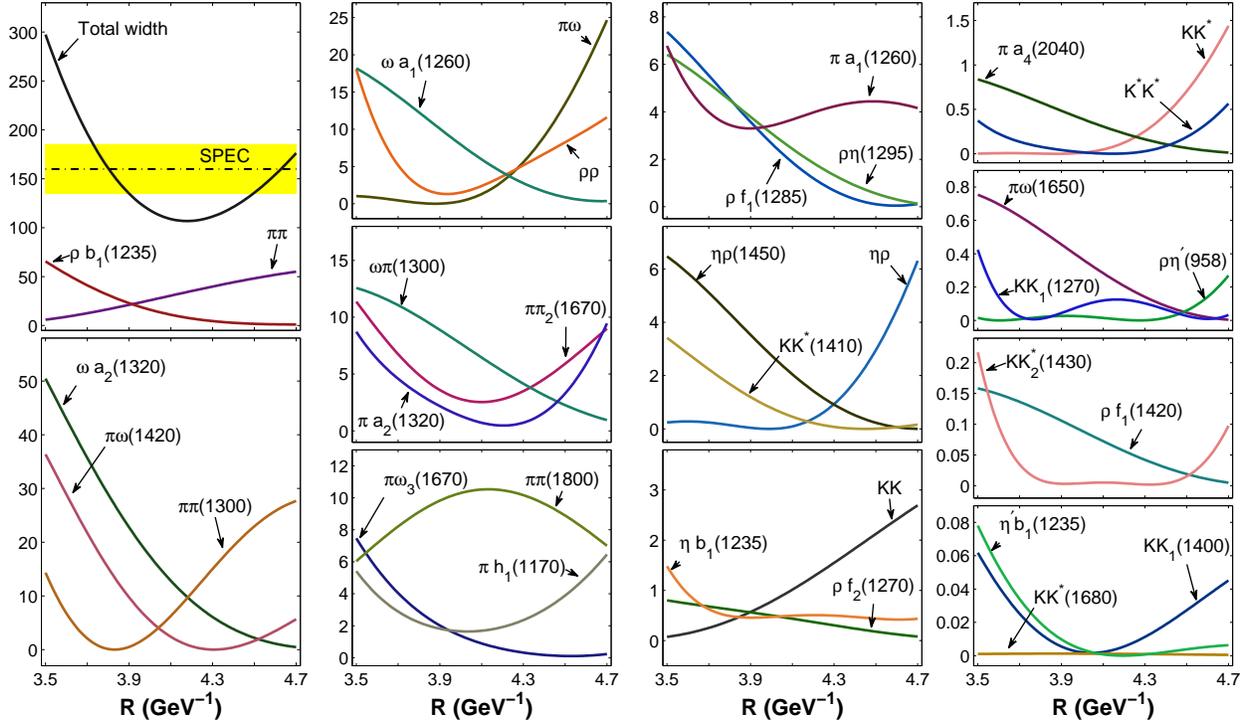}
\caption{(color online). The partial and total decay widths of $\rho_3(2250)$ dependent on the the $R$ value. Since the decay widths of the $\pi a_0(1450)$, $\rho\rho(1450)$, $K^*K_1(1270)$ and $\rho\eta(1475)$ channels are much smaller than that of the $K\bar K^*(1680)$ channel, we do not list these channels here. The dashed line with band is the experimental data \cite{Anisovich:2002su}.\label{3D3}}
\end{figure*}

In Fig. \ref{3D3} and Table \ref{ratio3}, the decay properties of $\rho_3(2250)$ as a $3^3D_3$ are illustrated. For higher $\rho_3$ meson, the decay behavior reflects the node effect, where $\rho_3(2250)$ decay widths are dependent on the $R$ value. If taking a typical value of $R=4.62$ GeV$^{-1}$, we can obtain the total decay width consistent with the experimental data \cite{Anisovich:2002su}. The corresponding main partial decay channels are $\pi\pi$, $\pi\pi(1300)$ and $\pi\omega$. Contrary to the former $\rho_3(1690)$ and $\rho_3(1900)$, the decay width of $\rho_3(2250)\to \rho\rho$ is small. At present, $\rho_3(2250)$ was observed in its $\pi\pi$, $K \bar K$, $\eta\pi\pi$, $\pi\omega$ and $\omega a_2(1320)$ decay channels.

\section{summary}\label{sec4}

In the past decades, many more $\rho/\rho_3$ states have been observed in experiments. How to categorize these $\rho/\rho_3$ states into the meson family is an intriguing research topic, which can improve our knowledge of light hadron spectrum. In this work, we systematically study the OZI-allowed two-body strong decay behaviors of the observed $\rho/\rho_3$ states, where the QPC model \cite{Micu:1968mk} is applied to the concrete calculation.

As shown in Fig. \ref{mass}, the mass spectrum analysis can provide preliminary information on these $\rho/\rho_3$ states, where their quantum numbers are assigned. Given these assignments, we perform the calculation of two-body strong decays of these $\rho/\rho_3$ states listed in Table \ref{exp-rho}. By comparing our theoretical results with the existing experimental data, the hadron structure properties of these $\rho/\rho_3$ states can be obtained and examined.

Besides getting the hadron structure properties of these $\rho/\rho_3$ states, our study also provides abundant decay information of these states, which can be as a valuable guidance to further experimental study on light hadron spectrum.

\vfil
\section*{Acknowledgement}

We would like to thank Qiang Zhao for useful discussions.
X.L. also would like to thank Takayuki Matsuki for reading our manuscript carefully. This project is supported by the National Natural Science
Foundation of China under Grants 11222547, 11175073, 11035006, the
Ministry of Education of China (FANEDD under Grant No. 200924,
SRFDP under Grant No. 20120211110002, NCET, the Fundamental
Research Funds for the Central Universities), the Fok Ying-Tong
Education Foundation (No. 131006).

\end{document}